\documentclass[a4paper,aps,prl,superscriptaddress,nofootinbib,nobibnotes,longbibliography,preprintnumbers,floatfix,twocolumn]{revtex4-2}
\usepackage[T1]{fontenc}
\usepackage[utf8]{inputenc}
\usepackage[english]{babel}
\usepackage[unicode, colorlinks=true, linkcolor=linkcolor, citecolor=linkcolor, filecolor=linkcolor, urlcolor=linkcolor, linktocpage, breaklinks]{hyperref}
\usepackage{csquotes,mathtools}
\usepackage[per-mode = symbol]{siunitx}
\usepackage{physics, tensor, bbm, mathrsfs}
\usepackage{amsmath, amssymb, amsfonts, amsthm}
\usepackage{graphicx, comment, wasysym, placeins}
\usepackage[dvipsnames, usenames]{xcolor}

\definecolor{linkcolor}{rgb}{0.0,0.3,0.5}
\definecolor{green}{rgb}{0.0,0.5,0.0}

\graphicspath{{./figures/}}

\newcommand{\theory}{\mathrm{x}}\newcommand{\textcomment}[1]{}

\begin{document}

\title{Ringdown Analysis of Rotating Black Holes\\ in Effective Field Theory Extensions of General Relativity}

\author{Simon Maenaut}
\affiliation{Institute for Theoretical Physics, KU Leuven, Celestijnenlaan 200D, B-3001 Leuven, Belgium}
\affiliation{Leuven Gravity Institute, KU Leuven, Celestijnenlaan 200D, B-3001 Leuven, Belgium}

\author{Gregorio Carullo}
\affiliation{Niels Bohr International Academy, Niels Bohr Institute, Blegdamsvej 17, 2100 Copenhagen, Denmark}
\affiliation{School of Physics and Astronomy and Institute for Gravitational Wave Astronomy, University of Birmingham, Edgbaston, Birmingham, B15 2TT, United Kingdom}

\author{Pablo A. Cano}
\affiliation{Departament de F\'isica Qu\`antica i Astrof\'isica, Institut de Ci\`encies del Cosmos\\
 Universitat de Barcelona, Mart\'i i Franqu\`es 1, E-08028 Barcelona, Spain}

\author{Anna Liu}
\affiliation{Department of Physics, The Chinese University of Hong Kong, Shatin, NT, Hong Kong}

\author{Vitor Cardoso}
\affiliation{Niels Bohr International Academy, Niels Bohr Institute, Blegdamsvej 17, 2100 Copenhagen, Denmark}
\affiliation{CENTRA, Departamento de F\'{\i}sica, Instituto Superior T\'ecnico -- IST, Universidade de Lisboa -- UL,
Avenida Rovisco Pais 1, 1049 Lisboa, Portugal}

\author{Thomas~Hertog}
\affiliation{Institute for Theoretical Physics, KU Leuven, Celestijnenlaan 200D, B-3001 Leuven, Belgium}
\affiliation{Leuven Gravity Institute, KU Leuven, Celestijnenlaan 200D, B-3001 Leuven, Belgium}

\author{Tjonnie G. F. Li}
\affiliation{Institute for Theoretical Physics, KU Leuven, Celestijnenlaan 200D, B-3001 Leuven, Belgium}
\affiliation{Leuven Gravity Institute, KU Leuven, Celestijnenlaan 200D, B-3001 Leuven, Belgium}
\affiliation{STADIUS Center for Dynamical Systems, Signal Processing and Data Analytics, KU Leuven, Kasteelpark Arenberg 10, B-3001 Leuven, Belgium}

\begin{abstract}
Quasinormal modes of rapidly rotating black holes were recently computed in a generic effective-field-theory extension of general relativity with higher-derivative corrections. 
We exploit this breakthrough to perform the most complete search for signatures of new physics in black hole spectra to date. 
We construct a template that describes the post-merger gravitational-wave emission in comparable-mass binary black hole mergers at current detector sensitivity, notably including isospectrality breaking. 
The analysis of all events with detectable quasinormal-driven ringdown signatures yields no evidence of higher-derivative corrections in the spectra, and we set an upper bound $\ell \lesssim \SI{35}{\kilo\metre}$ on the length scale of new physics.
Looking ahead, our scheme enables new studies on the capabilities of future detectors to robustly search for signatures of new gravitational physics.
\end{abstract}

\maketitle

\section{Introduction}
Gravitational wave (GW) detections from black hole (BH) binary coalescences by the LIGO-Virgo-KAGRA collaboration (LVK)~\cite{LIGOScientific:2014pky, VIRGO:2014yos, KAGRA:2013rdx, abbott2016observation, abbott2019gwtc1, LIGOScientific:2020ibl, KAGRA:2021vkt, LIGOScientific:2021sio} carry detailed information about the spacetime geometry near the event horizon. As such they provide a powerful observational window into strong and dynamical gravity, and therefore a promising route to search for possible signatures of new physics. 

Since the space of possible `beyond-GR' theories is large, most studies of new physics have relied on either agnostic frameworks~\cite{Agathos:2013upa,Meidam:2014jpa, LIGOScientific:2016lio,LIGOScientific:2020tif,LIGOScientific:2021sio,Dideron:2024xwm}, or theory-inspired schemes that encompass broad classes of theories~\cite{Yunes:2009ke,Tahura:2018zuq,Maselli:2019mjd,Carullo:2021dui,Maselli:2023khq,Payne:2024yhk}.
Recently, however, significant progress was made to compute precise strong-field signatures of specific beyond-GR theories motivated on theoretical principles.
These signatures came in the form of deviations in the post-Newtonian/Minkowskian~\cite{Sennett:2019bpc,AccettulliHuber:2020dal,Bernard:2022noq,Brandhuber:2024bnz,Brandhuber:2024qdn} or Effective-One-Body~\cite{Jain:2022nxs,Julie:2022qux} predictions for inspirals, and of robust numerical simulations of compact objects~\cite{Witek:2018dmd,Okounkova:2019dfo,Okounkova:2020rqw,East:2021bqk,East:2022rqi,Corman:2022xqg,Ripley:2022cdh,Evstafyeva:2022rve,AresteSalo:2022hua,AresteSalo:2023mmd,Cayuso:2023aht} modelling the merger phase (see e.g.~\cite{Julie:2024fwy}).

In parallel with these developments, much effort went into the extension of perturbation theory for rotating BHs beyond GR. The prime target on this front is the computation of perturbative corrections to the quasinormal mode (QNM) spectrum of BHs with significant angular momentum.
Since the QNMs ubiquitously drive the late-time relaxation, or ringdown, of the remnants of binary BH mergers~\cite{Berti:2009kk,Franchini:2023eda}, they are a key input in spectroscopic analyses of ringdown GW signals~\cite{Carullo:2019flw,Isi:2019aib,Finch:2021qph,Correia:2023ipz}.
Beyond-GR corrections to QNM frequencies of spherically symmetric stationary BHs were computed in~\cite{Cardoso:2009pk,Molina:2010fb,Blazquez-Salcedo:2016enn,Blazquez-Salcedo:2017txk,Cardoso:2018ptl,Tattersall:2018nve,McManus:2019ulj,Konoplya:2020bxa,deRham:2020ejn,Moura:2021nuh,Silva:2024ffz}. QNM shifts at linear or quadratic order in the rotation parameter were calculated in~\cite{Pani:2012bp,Pani:2013pma,Pierini:2021jxd,Wagle:2021tam,Srivastava:2021imr,Cano:2021myl,Pierini:2022eim} and exploited to analyse GW data in~\cite{Silva:2022srr}.
However, a change of paradigm was needed to accurately compute QNM shifts for the more rapidly rotating BHs that are of great astrophysical interest.
This came with the development of the Modified Teukolsky Equation~\cite{Li:2022pcy,Hussain:2022ins,Cano:2023tmv}, which made it possible to compute QNM spectral shifts up to very large order in the BH spin \cite{Cano:2023jbk,Cano:2024ezp}, marking a remarkable step forward in the prediction of these key observables\footnote{As an alternative approach, spectral techniques to compute QNMs beyond GR, specifically in Einstein-dilaton-Gauss-Bonnet (EDGB) gravity, were recently applied in Ref.~\cite{Miguel:2023rzp,Chung:2024ira,Chung:2024vaf,Blazquez-Salcedo:2024oek}.}.

When exploring extensions of GR, several conceptual complications arise from the vast landscape of theoretical possibilities and associated phenomenology.
This situation means that an effective-field-theory (EFT) approach is the optimal route to search for new physics signatures.
In this Letter, we adopt this approach, assuming that within the validity of the EFT the symmetries and degrees of freedom of GR are preserved.
Concretely, we consider the Einstein-Hilbert action with higher-curvature terms added, controlled by a length scale of new physics, as an EFT description of gravity at curvatures below this scale~\cite{Endlich:2017tqa,Cano:2019ore,deRham:2020ejn}.

That said, it has recently been questioned whether these EFTs can have observational consequences at all for GW astrophysics, mostly based on causality considerations \cite{Camanho:2014apa,deRham:2021bll,Serra:2022pzl,Caron-Huot:2022ugt,Kehagias:2024yyp}.
We take the view, however, that such causality-based arguments rather illustrate the difficulty to construct EFT models on the basis of our current fundamental principles. 
Viewed as classical theories, EFT extensions of GR with km-scale corrections are not ruled out by observations. GW observations offer a novel way to constrain the scales where beyond-GR corrections may enter. 
Therefore, it is important to flesh out the precise EFT predictions for GW signals, in order to make it possible to carry out such tests.
For an excellent summary of the solid motivations to test these theories in the dynamical and strong-field regime probed by LVK observations, see~\cite{Sennett:2019bpc}.

With this motivation in mind, we first use the corrections to the QNM spectrum of rapidly-rotating BHs calculated in \cite{Cano:2023jbk,Cano:2024ezp} for this class of theories, to model the ringdown signal and constrain these theories.
With these predictions at hand, we then analyse all recently observed events in the GWTC-3 catalogue~\cite{LIGOScientific:2020ibl, KAGRA:2021vkt}, with clear and confident ringdown signatures~\cite{LIGOScientific:2020tif,LIGOScientific:2021sio}.
This completes the first spectroscopic analysis of GW data from binary coalescences in a beyond-GR theoretical frame that robustly models the high-rotation rate of astrophysical BHs. 
As such, it constitutes a critical milestone in high-precision searches for new physics potentially hidden in the GW patterns generated in binary BH mergers.

\section{Higher-derivative gravity}

A general EFT extension of GR constructed purely from Riemann tensor contractions that preserves parity symmetry to eight derivatives is given by the action\footnote{We do not include any quadratic curvature terms, since these do not modify the vacuum BH solutions from Einstein Gravity and only introduce additional separate branches of BH solutions.} \cite{Cano:2019ore}
\begin{align}\label{eq:EFT}
S=\int d^4x\frac{\sqrt{|g|}}{16\pi}\bigg[R+\ell^4
\lambda_{\rm ev}\mathcal{R}^3
+\ell^6\left(\lambda_{1}\mathcal{C}^2+\lambda_{2}\tilde{\mathcal{C}}^2 \right)
\bigg]
\end{align}
This includes the higher-derivative curvature scalars
$ \mathcal{R}^3 =\tensor{R}{_{\mu\nu }^{\rho\sigma}}\tensor{R}{_{\rho\sigma }^{\delta\gamma }}\tensor{R}{_{\delta\gamma }^{\mu\nu }}$
together with 
$\mathcal{C} =R_{\mu\nu\rho\sigma} R^{\mu\nu\rho\sigma}$ and
$\tilde{\mathcal{C}} =R_{\mu\nu\rho\sigma} \tilde{R}^{\mu\nu\rho\sigma}$,
where $\tilde{R}_{\mu\nu\rho\sigma}=\frac{1}{2}\epsilon_{\mu\nu\alpha\beta}\tensor{R}{^{\alpha\beta}_{\rho\sigma}}$ is the dual Riemann tensor. 
There is a new length scale $\ell$ related to the cut-off of the EFT, $\ell \sim 1/\Lambda_\mathrm{cutoff}$, while the coefficients $\lambda_{\text{ev, 1, 2}}$ are dimensionless. With the physical scale of the system set by a BH of mass $M$, we introduce the dimensionless couplings:
\begin{align}\label{eq:alpha}
\alpha_\mathrm{ev}&=\frac{\ell^4\lambda_\mathrm{ev}}{M^4}\, ,& \alpha_{1}&=\frac{\ell^6\lambda_{1}}{M^6}\, ,& \alpha_{2}&=\frac{\ell^6\lambda_{2}}{M^6}\, .
\end{align}
These characterise the size of the relative corrections to GR, which thus will become larger for smaller BHs.
We assume $\abs{\alpha_{\theory}} \ll 1$ throughout the analysis, where $\theory \in \{\mathrm{ev}, 1, 2\}$ so that we are well within the EFT regime and can apply an expansion in these couplings. 
We do not impose any further constraints in the present analysis.

The rotating BH solutions of (\ref{eq:EFT}) take the form of a modified Kerr metric that can be found analytically as a power series in the dimensionless spin $\chi=a/M$ (where $a=J/M$ is the specific angular momentum) using the methods of \cite{Cano:2019ore}. We refer to that work for details; here we simply remark that this technique allows one to describe BHs of large angular momentum by carrying out the spin expansion to high order \cite{Cano:2023qqm}.

\section{Quasinormal mode spectrum}

Gravitational perturbations of rotating BHs in the theories of (\ref{eq:EFT}) can be described in terms of the four perturbed Weyl scalars \textcomment{These conjugate variables $\delta\Psi_{0,4}^{*}$ are considered independent variables, since we consider a complex metric perturbation. Hence, the symbol $*$ denotes Newman-Penrose (NP) conjugation, obtained upon the exchange $m_{\mu}\leftrightarrow \bar{m}_{\mu}$ in the NP frame. For a complex metric perturbation, this is no longer equivalent to complex conjugation.} $\delta\Psi_{0}$, $\delta\Psi_{4}$, $\delta\Psi_{0}^{*}$ and $\delta\Psi_{4}^{*}$ using the framework of the Modified Teukolsky Equations \cite{Li:2022pcy,Hussain:2022ins,Cano:2023tmv}.
As shown in \cite{Cano:2023tmv} and discussed in \cite{Cano:2023jbk,Cano:2024ezp}, it is possible to reduce the Modified Teukolsky equations for a theory close to GR to a system of four radial equations for four master variables $R^{lm}_{s}(r)$, $R^{*lm}_{s}(r)$ by decomposing the Weyl scalars in the spin-weighted spheroidal harmonics.
Here, $(l,m)$ are the angular mode harmonic numbers and $s=\pm 2$ is the spin weight of the variable.
At first order in the coupling of higher-derivative corrections ($\alpha_\mathrm{x}$), these variables satisfy a decoupled radial Teukolsky equation with a perturbed potential:
\begin{align}\label{eq:correctedradial}
\Delta^{-s+1}\frac{d}{dr}\left[\Delta^{s+1}\frac{dR}{dr}\right]+\left(V+ \alpha_\theory \, \delta V_\theory\right) R=0\, ,
\end{align}
where we have suppressed the labels of $(s,l,m)$ indices.

Quasinormal modes are the solutions of the equations (\ref{eq:correctedradial}) with outgoing boundary conditions at infinity and the horizon.
They capture how a BH relaxes after a local perturbation.
Eq.(\ref{eq:correctedradial}) can be solved as an eigenvalue problem for the frequency $\omega$, which yields the spectrum of QNM frequencies.
For higher-derivative gravity theories, the Teukolsky equations depend on the polarization of the perturbation.
In parity-preserving theories, such as those that we consider, modes of odd and even parity decouple.
The QNM frequencies $\omega$ are then obtained by solving any of the radial equations.

For QNM frequencies, accounting for linear corrections with respect to the Kerr values $\omega^{(0)}$, we write
\begin{align}\label{eq:deltaomegaeven}
\omega^{\pm}_{\theory,lmn}=\omega^{(0)}_{lmn}+
\alpha^{\phantom{\pm}}_{\theory} \delta\omega^{\pm}_{\theory,lmn} ,
\end{align}  
where the label $\pm$ refers to the polarisation. Generically we observe that $\delta\omega^{+}_{lmn}\neq \delta\omega^{-}_{lmn}$, and isospectrality is broken. 
We refer the reader to \cite{Cano:2023jbk,Cano:2024ezp} for details on how the shifts of QNM frequencies can be calculated.

We use the results of a polynomial fit as a function of the dimensionless spin $\chi$ to model the linear order corrections to the QNM frequencies for each polarisation and theory,
\begin{align}\label{eq:deltaom}
\delta \omega^{\pm}_{\theory,lmn} = \sum_{k=0}^{K} c^{(k)\pm}_{\theory,lmn} \, \chi^k \, ,
\end{align}
where the coefficients for the dominant fundamental modes and their overtones were calculated for $K = 12 $. These polynomials provide a good approximation to the QNM shifts up to the desired value of $\chi=0.7$ and even for $\chi\sim 0.8$ \cite{Cano:2023jbk,Cano:2024ezp}. We have verified with different numerical techniques that an increase of the order of the spin expansion in these theoretical calculations confirms the values of coefficients and only changes the highest coefficient significantly.

\section{Template construction}

To construct the waveform template used to analyse the data, we rely on the \texttt{pyRing} python package, which is tailored to model and analyse ringdown signals~\cite{pyRing,Carullo:2019flw}.
Although a diverse set of GR and beyond GR models are implemented in \texttt{pyRing}, we found it more convenient to introduce a new parametrisation for the effect of the higher-derivative corrections on the ringdown waveform. 
We start from the Kerr model and separate the metric polarisations as these are no longer isospectral \textcomment{need single number for entire set of equations}
\begin{align}
&\tensor{h}{_{+}} - i h_{\cross} = \frac{M_\mathrm{obs}}{D_L} \sum_{l=2}^{\infty} \sum_{m=-l}^{+l} \sum_{n=0}^{\infty} \sum_{q=\pm} \left( h^{P,q}_{lmn} + h^{N,q}_{lmn} \right),\notag\\
&h^{P,q}_{lmn} = \mathcal{A}^{P,q}_{lmn} \, \mathcal{S}_{lmn}  (\iota, \varphi) \, e^{+i \, (t-t_{s}) \, \omega^{q\phantom{*}}_{lmn} } \, \theta(t-t_{\rm s}),\notag\\
&h^{N,q}_{lmn} = \mathcal{A}^{N,q}_{lmn} \, \mathcal{S}_{lmn}  (\iota, \varphi) \, e^{-i \, (t-t_{s}) \, \omega^{q*}_{lmn} } \, \theta(t-t_{\rm s})
\end{align}
where $\omega^q_{lmn}$ is the modified QNM frequency, expressed as a function of the red-shifted mass of the remnant BH as observed in the detector frame $M_\mathrm{obs}$, and its spin $a$.
The angle $\iota$ denotes the inclination of the rotation axis of the final BH relative to the observer’s line of sight and $\varphi$ the azimuthal angle of the line of sight in the BH frame. 
Here $\mathcal{S}_{lmn}$ are spin-weighted spheroidal harmonics, and $t_{\rm s} = t_0$ is the start time of the template.

Each QNM labelled by the indices $lmn$ has two complex frequencies $\omega^\pm_{lmn}$ and a priori also two unknown complex amplitudes $\mathcal{A}^{P/N,\pm}_{lmn}$, where the contributions with a positive and negative real part of the frequency for each QNM are labelled with $P$ and $N$ respectively.
Assuming the spins of the binary progenitors were aligned with the orbital angular momentum, reflection symmetry implies $\mathcal{A}^{ N,\pm}_{lmn} =  (-1)^l \, ( \mathcal{A}^{P,\pm}_{lmn} )^*$, which halves the number of free parameters per mode.
For theories that preserve parity symmetry, as the ones considered in the analysis, this is a very good approximation for the current dataset for the purpose of searching for new physics.
Here, $m < 0$ contributions represent counter-rotating modes~\cite{Berti:2005ys,Lim:2019xrb,Li:2021wgz}, which are only significantly excited in systems with large progenitors spins anti-aligned with the orbital angular momentum, hence negligible for the current dataset.

To model the modified complex QNM frequency $\omega_{lmn}$ for a specific higher-derivative correction, we use \eqref{eq:deltaomegaeven} and \eqref{eq:deltaom}. The theory-dependent coefficients $c^{(j)\pm}_{\theory,lmn}$ are made available within \texttt{pyRing}~\cite{pyRing}. In what follows we remove the theory index $\theory$ to ease the notation.

Assuming that we have a single correction in the action \eqref{eq:EFT}, we can reabsorb the corresponding coupling $\lambda$ into the length scale of the EFT $\ell$, so we can always set $\lambda=\pm 1$. From \eqref{eq:alpha}, we express $\alpha$ as
\begin{align}
\alpha &= \lambda \left(\frac{\ell}{M} \right)^p
= \operatorname{sign}(\ell)\left( 
\frac{\abs{\ell} \cdot (1+z)}{M_\text{obs}}\right)^p\, ,
\end{align}
where $p$ is the power of the scaling of the coupling, equal to $4$ for $\lambda=\lambda_\mathrm{ev}$ and $6$ for $\lambda=\lambda_{1,2}$, and we allow $\ell$ to be negative to account for the sign of $\lambda$. 
Since the length scale modifying the BH dynamics is related to the source dynamics, the mass scale of the BH in the rest frame $M$ needs to be scaled with the redshift factor compared to the mass observed in the detector $M_\text{obs} = M \left( 1+z \right)$.
This signals a breaking of GR scale invariance.
If any EFT modification would be detected and consequently a specific value of $\ell$ inferred, it would then be possible to measure the source-frame mass $M$ with subsequent observations.
This would allow us to perform measurements of the redshift, hence the cosmological evolution, entirely using GW signals from BH binaries.

\begin{figure}\centering
\includegraphics[width=\linewidth]{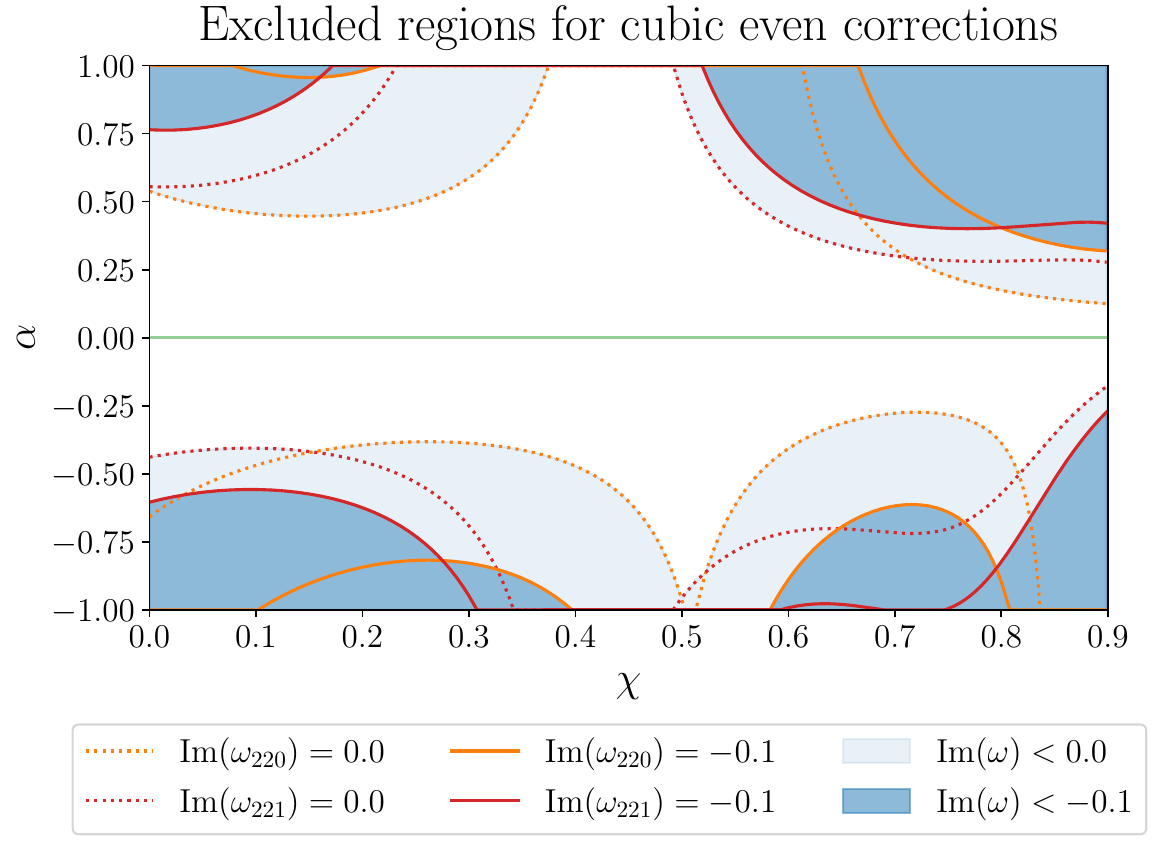}
\caption{Parameter space of effective-field-theory corrections to the general relativistic quasinormal modes spectrum, as a function of coupling constant and black hole rotational rate.
Regions where the fundamental and first overtone $(l,m,n) = (2,2,0), (2,2,1)$ modes appear to grow exponentially $\Im[\omega] < 0$ (dotted lines), or grow faster than $\Im[\omega] < -0.1$ (solid lines) are highlighted, for cubic even corrections.}
\label{fig:limits}
\end{figure}

A limitation of only knowing the QNMs spectrum to linear order in the coupling is that it is still possible, even with $\alpha < 1$, to have exponentially growing modes for certain intervals of $\chi$, dependent on the polarisation and theory.
These features are expected to be removed when including higher-order corrections, and signal a breaking of the linear approximation.
Hence, in our analysis, apart from limiting the correction factor to $\abs{\alpha} < 1$, we also exclude regions of the parameter-space with modes that grow exponentially.
Apart from being unphysical, this behaviour also causes numerical issues when calculating the likelihood when modes grow quickly ($\Im[\omega] < -0.1$). 
Hence, we exclude these regions a priori, instead of letting the data disfavour the regions of the parameter space where it happens, which would happen anyway given the damped morphology of the GW signal.
Fig.~\ref{fig:limits} illustrates the location of these effects, with the constraint on the parameter space added as a function of $\alpha$ and $\chi$.

The spin of the final remnant BH was limited to $\chi \leq 0.93$ in the analysis, since the uncertainty on the polynomial fit increases rapidly near extremal spin, while the magnitude of the linear corrections to most of the QNM frequencies seem to grow significantly. 
This limit does not affect the obtained constraints, since no near-extremal remnant spins have yet been observed.

Finally, in the analysis we searched for the dominant $lmn = 220$ fundamental mode and also its first overtone $lmn = 221$.  
Although the addition of overtones with constant-amplitudes near the waveform peak~\cite{Giesler:2019uxc, Isi:2019aib} does not correspond to physical BH vibrational spectra~\cite{Baibhav:2023clw} (see also ~\cite{Nee:2023osy,Zhu:2023mzv} for wave packets scattering), their phenomenological usage significantly boosts the recovered signal-to-noise ratio, even beyond-GR~\cite{Cayuso:2023aht}.
We do not include other modes, since no higher angular mode has so far been confidently detected~\cite{LIGOScientific:2021sio,Siegel:2023lxl,Gennari:2023gmx}.

\section{Results and discussion}

\begin{figure}\centering
\includegraphics[width=\linewidth]{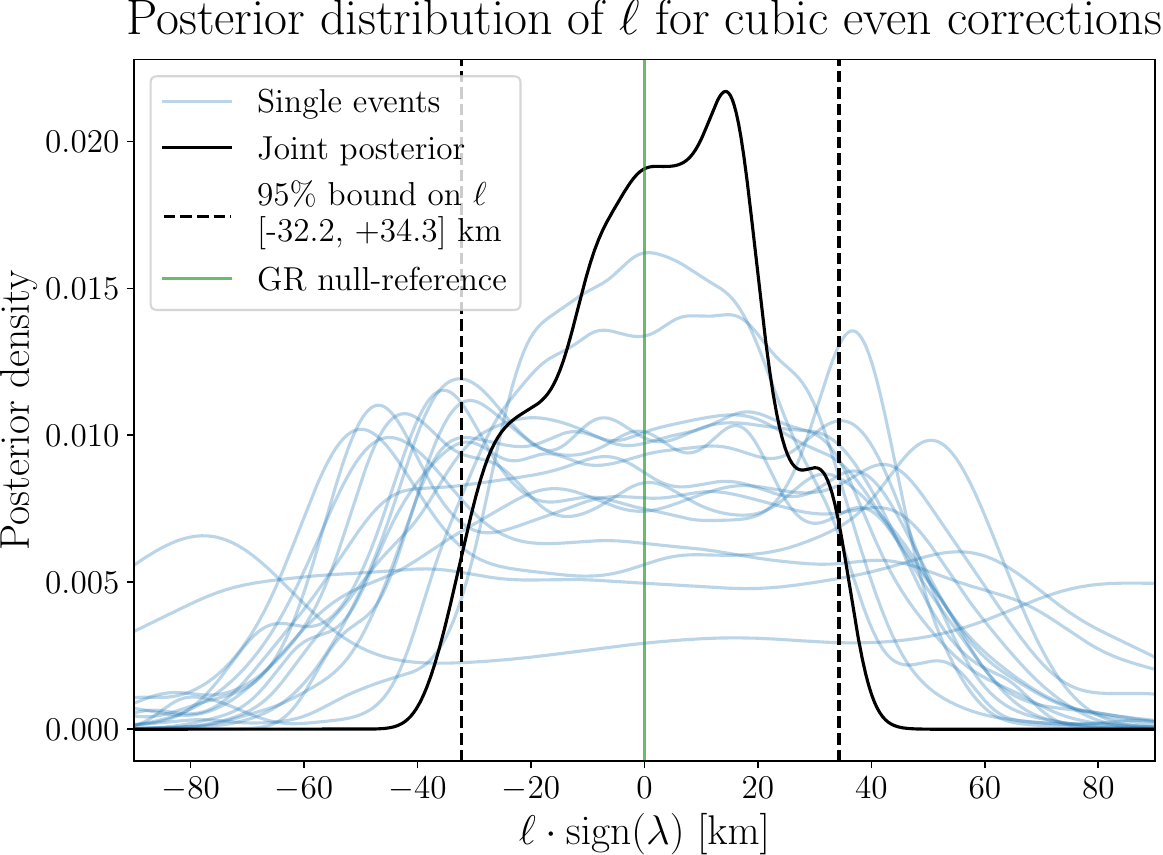}
\caption{Posterior probability distributions of the length scale of new physics, $\ell$, appearing in the effective-field-theory action (cubic even case).
Constraints are derived from the time-domain spectroscopic analysis with \texttt{pyRing} of post-merger signals from binary black hole coalescences, contained in the GWTC-3 catalogue from the LVK collaboration.}
\label{fig:posterior}
\end{figure}

We now analyse the observed GW events from the BH mergers contained in the GWTC-3 catalogue~\cite{LIGOScientific:2020ibl,KAGRA:2021vkt} that present detectable ringdown signatures. For this we make use of the ringdown analysis python code \texttt{pyRing} \cite{pyRing,Carullo:2019flw}, specifically written for the analysis of the post-merger phase signals with a time-domain likelihood formulation.
Such framework has been used within all ringdown-only spectroscopic analyses from the LVK collaboration in the GWTC-2~\cite{LIGOScientific:2020iuh,LIGOScientific:2020ufj,LIGOScientific:2020tif} and GWTC-3~\cite{LIGOScientific:2021sio} catalogues, and is thus extensively tested.
We use the same event-selection criteria, and identical settings to the latter analyses also with regard to the data conditioning, the sampling and the likelihood parameters.
We do not include in our dataset the event GW191109\_010717 because of data-quality issues~\cite{LIGOScientific:2021sio}.
To increase slightly the amount of signal-to-noise ratio (SNR) captured~\cite{Isi:2023nif,Siegel:2024jqd}, at the expense of an increase in computational cost compared to the LVK analyses, we increased the analysis segment to~$\SI{0.2}{\second}$.
We isolate the GW signal portion that follows the peak of the reconstructed $h_{+}^2 + h_{\times}^2$ and fix $t_{\rm start}$ to this peak.
Following standard practise in spectroscopic analyses, we assume the maximum likelihood sky position extracted from a full signal template to propagate this estimate to multiple detectors~\cite{Isi:2021iql,LIGOScientific:2020ibl}.
This is allowed since GW propagation is not modified in our setup.
We apply uniform priors on all the sampled parameters: the remnant BH redshifted mass $M_\text{obs} \in [10,500] \, M_{\odot}$, remnant spin $\chi \in [0.00,0.93]$, the QNM amplitudes $A_{lmn} \in [0, 50]$, phases $\phi_{lmn} \in [0, 2\pi]$ and EFT scale $\ell \in [-740, +740]\, \si{\km}$ (including $\operatorname{sign}(\lambda)$). 
To avoid prior effects having a sizeable impact on the $\ell$ estimate, the luminosity distance is also sampled under a uniform prior, and constrained within the $95 \%$ bound obtained from GWTC-3 analyses.
We repeat the analysis both when assuming QNM complex frequencies to be dictated by GR (in which case we fix $\ell=\SI{0}{\km}$), and by the EFT corrections under consideration, allowing us to construct an estimate of the Bayesian evidence for the presence of any deviations in the Einstein-Hilbert action.
For all the events analysed, we verified that the range of $\ell$ (in km) supported by its marginalised posterior was at a clearly lower scale than the range of $M$ supported by its marginalised posterior (converted from $M_\odot$ to km in geometric units and taking into account the red-shift). A significant overlap between these posteriors would break the EFT assumption.
There is one event, GW190708\_232457, where the BH mass posterior is so broad that the $\ell$ range becomes comparable to the EFT cutoff, hence the results do not satisfy the EFT assumption (see e.g. Sec.IV.D of~\cite{Silva:2022srr}).
This event was therefore excluded from further analysis.

We report the marginalised posterior distributions of the $\ell$ parameter in Fig.~\ref{fig:posterior}, for all events and the cubic even corrections (the posteriors of the quartic corrections, with similar qualitative results, are in the ancillary materials).\\
Posterior distributions for all events are centred around zero, with various degree of spread due to the difference in SNR among different events, together with a roughly corresponding degree of noisy features.
Since $\ell$ is fixed by a universal fundamental constant that is common to all events, combining the single events constraints into a unique posterior probability distribution can be simply achieved by first constructing a standard kernel density estimation, and then multiplying the likelihood functions (under our assumption of a flat prior)~\cite{Zimmerman:2019wzo}.
The result of this is shown in Fig.~\ref{fig:posterior} (solid black line). 
As expected, the combined result is less sensitive to noise-induced features of single events (light-blue lines).
We note that although the shape of the combined result can be quite sensitive to individual results, the constraints that come from it remain robust even if individual events are removed.

Posterior distributions of remnant BH parameters and complex amplitudes are consistent between the GR and EFT analyses, with an expected broadening in the EFT case due to the increase in the number of parameters.
The agreement of these two sets of posteriors, together with the large support for $\ell=0$ shown in Fig.~\ref{fig:posterior} indicates that no sign for violations of GR is present in the dataset.
This result is made quantitative in Table~\ref{tab:constraints}, reporting the ratio of the global Bayesian evidences comparing the GR and EFT hypotheses, namely the Bayes Factors.
We find no statistically significant evidence for a violation of GR.\\
These constraints improve those obtained in Ref.~\cite{Silva:2022srr}, which used a \texttt{pSEOBNR} waveform model that modified the ringdown to include corrections to the dominant QNM frequency and damping time to linear order in the spin.

\begin{table} 
\begin{tabular*}{\linewidth}{@{\extracolsep{\fill}}lcr}
\hline
Theory & $\operatorname{sign}(\lambda) \cdot \ell$ in \si{\kilo\meter}& $\ln(\mathcal{B}_\mathrm{EFT}/\mathcal{B}_\mathrm{GR})$\\
\hline
Cubic even& $[-32.2,\, +34.3]$ & $[-2.0,\, +1.0]$ \\[\baselineskip]
Quartic 1 & $[-24.9,\, +35.0]$ & $[-2.1,\, +1.4]$ \\[\baselineskip]
Quartic 2 & $[-27.0,\, +38.7]$ & $[-1.7,\, +0.9]$ \\[\baselineskip]
\hline
\end{tabular*}
\caption{Summary of the constraints obtained at a 95~\% level for each EFT correction, by combining posterior distributions of single events from GWTC-3, and the range of Bayes factors obtained from all events when comparing models.}
\label{tab:constraints}
\end{table}

\section{Conclusions}

We presented the first time-domain spectroscopic search valid up to high BH rotation rates in a symmetry-preserving EFT extension of GR, constructed under minimal assumptions.
Our study focused on the post-merger phase of binary BH mergers and relied crucially on the recent accurate predictions of the QNM frequency shifts in these theories.
We find no evidence for new physics signatures in the quasinormal modes of the observed GWs from the BH mergers in the GWTC-3 catalogue. 
The logarithm of the Bayes factor [$\ln(\mathcal{B}_\text{EFT}/\mathcal{B}_\text{GR})$] of each EFT model with respect to GR was less than 1.5 for all events and negative for most events, compatible with statistical fluctuations~\cite{LIGOScientific:2020tif}.
Combining the information from all events we obtain a constraint on the length scale associated with each of the EFT corrections.
These constraints are $\ell < \SI{34.3}{\kilo\metre}$ for the cubic correction, $\ell < \SI{35.0}{\kilo\metre}$ for quartic 1, and $\ell < \SI{38.7}{\kilo\metre}$ for quartic 2.  

Our analysis can readily be extended to EFTs with parity-violating terms.
We leave such extension to future study, except to note that the assumption of symmetry that couples the amplitudes of modes would no longer be correct, implying a doubling of the number of free amplitude parameters and hence an increased computational cost. 
For other theory-specific corrections, incl. EDGB, a similar analysis should be possible if leading corrections to the fundamental modes and the first overtones of QNM frequencies are known for sufficiently high rotation.

The next-generation ground-based GW detectors such as the Einstein Telescope and Cosmic Explorer~\cite{Punturo:2010zza,Reitze:2019iox} will enable the detection of BH mergers with much larger SNR. Moreover, given their improved sensitivity at higher-frequencies, they will make it possible to observe the ringdown for smaller black holes. Since these have horizon regions with larger spacetime curvature, this will dramatically boost our capability to search for higher-derivative corrections to GR.
The expected sensitivity of these detectors to theory-specific corrections to the ringdown signal was studied in Ref.~\cite{Maselli:2023khq}, and it would be interesting to repeat a similar analysis in our setup.
One way to improve these constraints might be through modelling the amplitudes of the QNMs for rotating BHs in these theories, either using numerical relativity simulations or from pre-merger predictions from parametrized-post-Newtonian or parametrized-post-Einsteinian frameworks.
This method would allow one to extract significantly more information from the ringdown waveform, enabling even higher sensitivity.
Overall, theory-specific predictions of quasinormal spectra for rotating BHs will be essential to exploit the full potential of next-generation GW detectors in searching for new physics in the high-curvature regime of gravity. Our results constitute a stepping-stone in this direction.

\section*{Acknowledgements}

\begin{acknowledgments}

We are grateful to Emanuele Berti, Giada Caneva, Lodovico Capuano, Adrian Chung, Poul Damgaard, Guillaume Dideron, Nicola Franchini, Vasco Gennari, Luis Lehner and Sebastian V\"olkel for stimulating discussions and valuable comments.
We thank Will Farr, Maximiliano Isi, Harrison Siegel and Hector Silva for comments on the draft.\\
\noindent\textit{Software --}
Contents of this manuscript were derived using the publicly available \texttt{python} software packages: \texttt{corner}, \texttt{cpnest}, \texttt{cython}, \texttt{gwpy}, \texttt{lalsuite}, \texttt{matplotlib},  \texttt{numpy}, \texttt{pandas}, \texttt{pyRing}, \texttt{scipy} and \texttt{seaborn} ~\cite{corner,cpnest,cython, gwpy,lalsuite,matplotlib,numpy,pandas,scipy,seaborn,pyRing}.
The \texttt{pyRing} commit is \texttt{6553253b} of the \texttt{EFT\_QNMs} branch, built on top of version \texttt{v2.6.0}.

\noindent\textit{Funding --} 
The work of P. A. C. received the support of a fellowship
from ``la Caixa'' Foundation (ID 100010434) with code
LCF/BQ/PI23/11970032.
S.M. and T.H. acknowledge support from the Flemish inter-university project IBOF/21/084.
S.M. would like to thank the L\'eon Rosenfeld Foundation for supporting his research visit to the Strong group at the Niels Bohr Institute during the final stages of this work.
G.C. acknowledges funding from the European Union’s Horizon 2020 research and innovation program under the Marie Sklodowska-Curie grant agreement No. 847523 ‘INTERACTIONS’.
G.C. and V.C. acknowledge support from the Villum Investigator program by the VILLUM Foundation (grant no. VIL37766) and the DNRF Chair program (grant no. DNRF162) by the Danish National Research Foundation.
This project has received funding from the European Union's Horizon 2020 research and innovation programme under the Marie Sklodowska-Curie grant agreement No 101131233.
This material is based upon work supported by NSF's LIGO Laboratory which is a major facility fully funded by the National Science Foundation.
The authors are grateful for computational resources provided by the LIGO Laboratory, supported by National Science Foundation Grants: PHY-0757058 and PHY-0823459.
This research has made use of data or software obtained from the Gravitational Wave Open Science Center (gwosc.org), a service of the LIGO Scientific Collaboration, the Virgo Collaboration, and KAGRA. This material is based upon work supported by NSF's LIGO Laboratory which is a major facility fully funded by the National Science Foundation, as well as the Science and Technology Facilities Council (STFC) of the United Kingdom, the Max-Planck-Society (MPS), and the State of Niedersachsen/Germany for support of the construction of Advanced LIGO and construction and operation of the GEO600 detector. Additional support for Advanced LIGO was provided by the Australian Research Council. Virgo is funded, through the European Gravitational Observatory (EGO), by the French Centre National de Recherche Scientifique (CNRS), the Italian Istituto Nazionale di Fisica Nucleare (INFN) and the Dutch Nikhef, with contributions by institutions from Belgium, Germany, Greece, Hungary, Ireland, Japan, Monaco, Poland, Portugal, Spain. KAGRA is supported by Ministry of Education, Culture, Sports, Science and Technology (MEXT), Japan Society for the Promotion of Science (JSPS) in Japan; National Research Foundation (NRF) and Ministry of Science and ICT (MSIT) in Korea; Academia Sinica (AS) and National Science and Technology Council (NSTC) in Taiwan

\end{acknowledgments}

\bibliography{references}

\onecolumngrid

\newpage

\appendix*

\section{Results for quartic theories}
\begin{figure}[h]\centering
\includegraphics[width=0.48\linewidth]{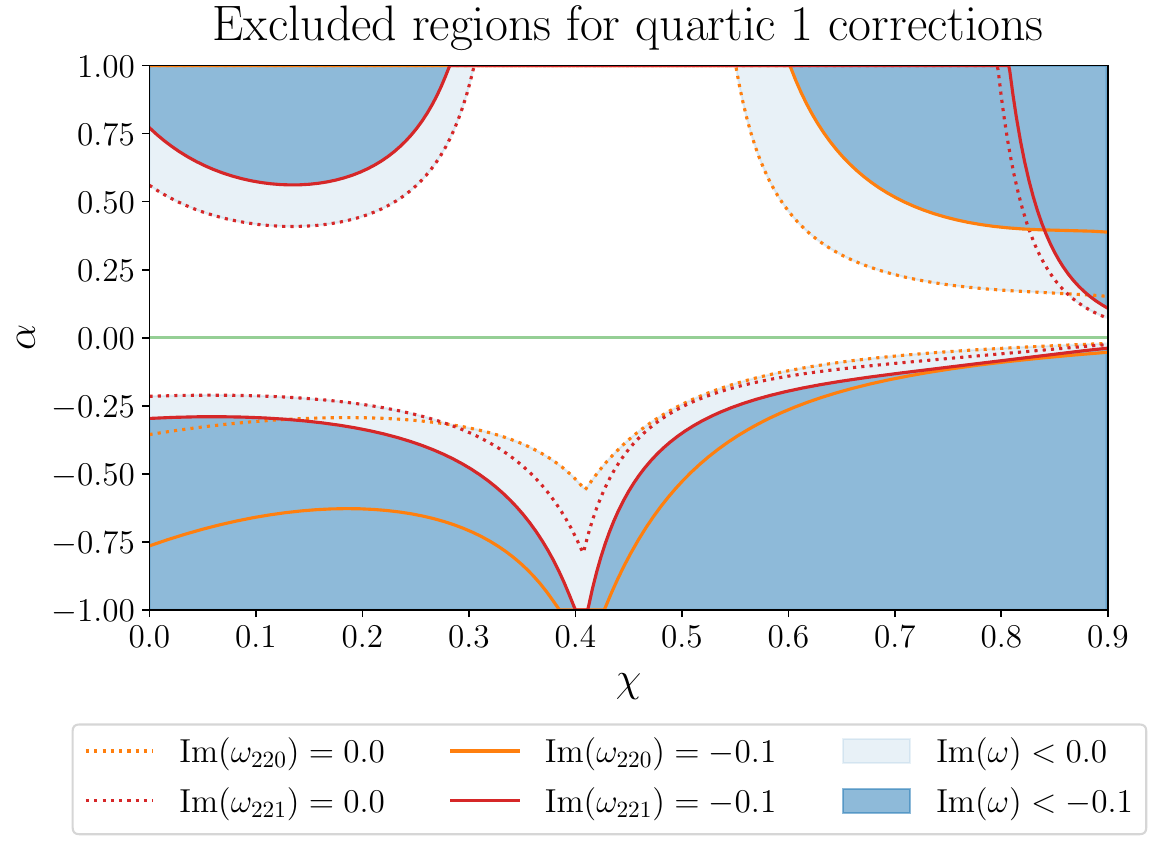}\hfill
\includegraphics[width=0.48\linewidth]{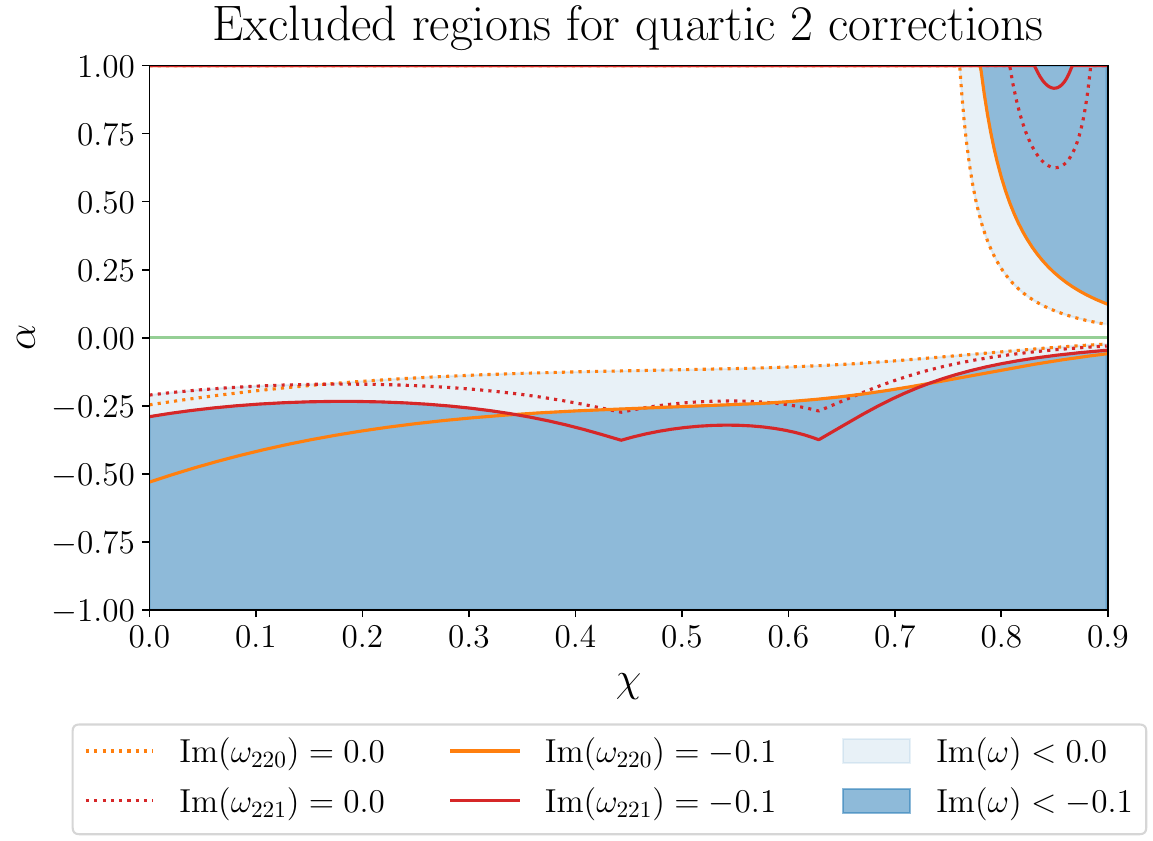}
\caption{Regions where $(2,2,0)$ and $(2,2,1)$ modes appear to grow exponentially $\Im[\omega] < 0$ (dotted lines) for quartic 1 and 2 corrections, and grow faster than $\Im[\omega] < -0.1$ (solid lines).}
\label{fig:appendix-regions}
\end{figure}
\begin{figure}[h]\centering
\includegraphics[width=0.48\linewidth]{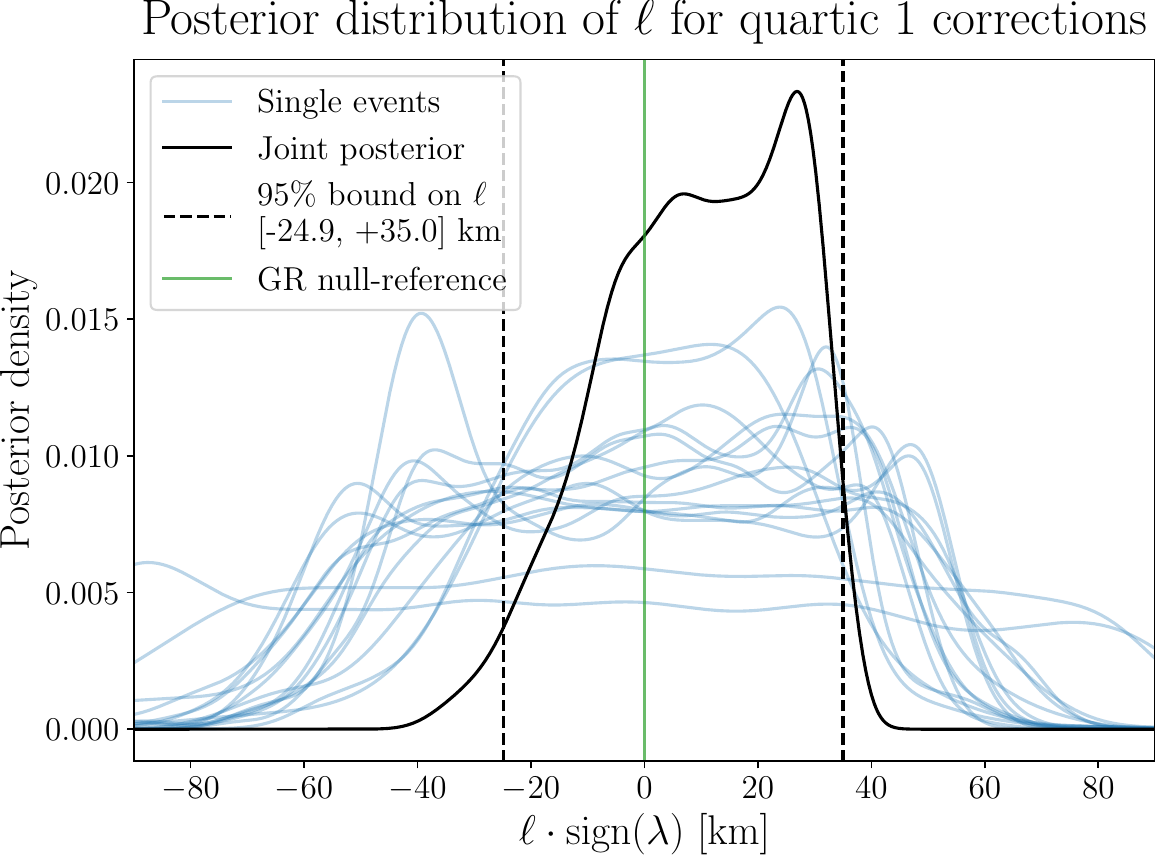}\hfill
\includegraphics[width=0.48\linewidth]{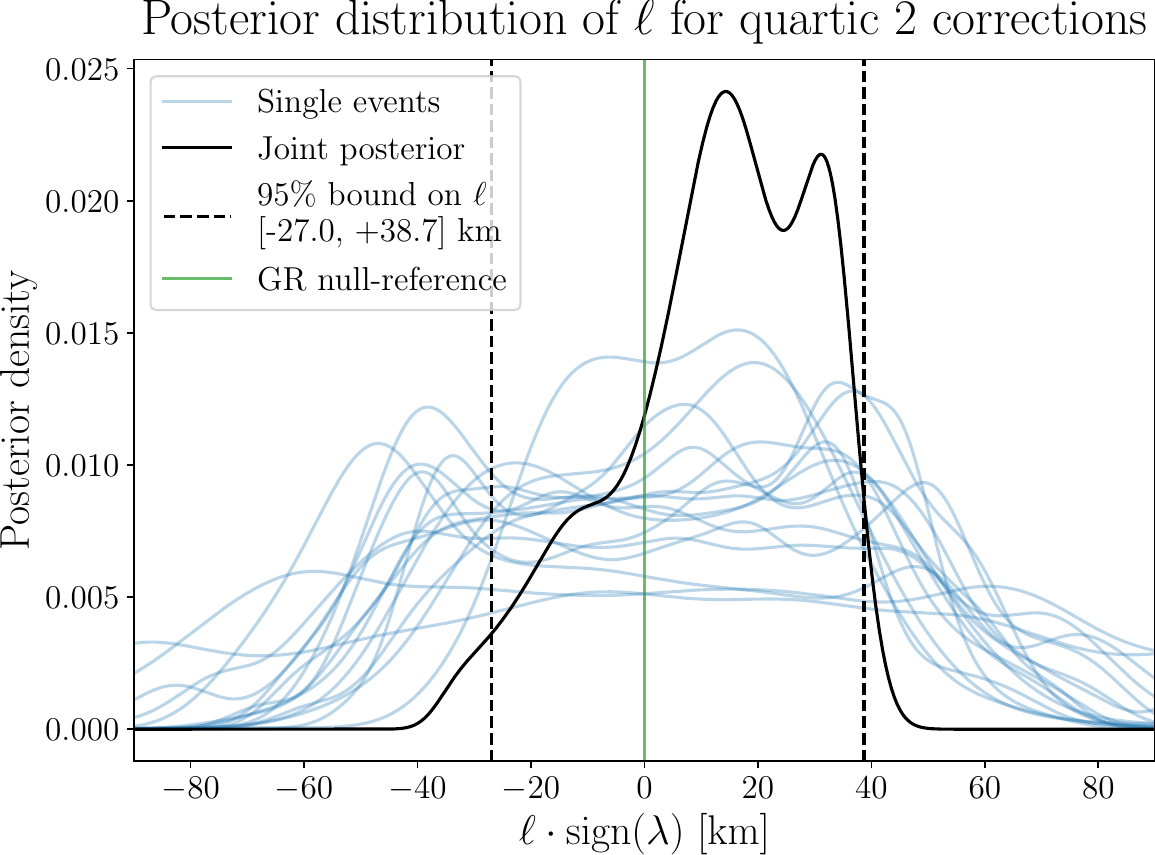}
\caption{Posterior distributions of $\ell$ for quartic 1 and 2 corrections obtained from the analysis of the GWTC-3 catalogue.}
\label{fig:appendix-posterior}
\end{figure}
\end{document}